\documentclass[11pt]{article}
\usepackage[latin1]{inputenc}
\usepackage[english]{babel}
\usepackage[namelimits]{amsmath}
\usepackage{amssymb}
\usepackage{amsmath}
\usepackage{amsthm}
\usepackage{epsfig,multicol,bbm}

\begin{document}
\title{Graceful exit from inflation to radiation era with rapidly decreasing agegraphic potentials}
\author{Stefano Viaggiu
\\
Dipartimento di Matematica, Universit\'a "Tor Vergata" Roma,\\
Via della Ricerca Scientifica, 1\\
Rome, Italy 00133,\\
viaggiu@axp.mat.uniroma2.it\\
Marco Montuori
\\
SMC-ISC-CNR and Dipartimento di Fisica,\\
Universit\'a "La Sapienza" Roma,\\
Ple. Aldo Moro 2\\
00185, Rome, Italy,\\
marco.montuori@roma1.infn.it,\\
}
\date{\today}\maketitle
\begin{abstract}
We present a class of models where both the primordial inflation and the late times de Sitter phase are driven by  
simple phenomenological agegraphic potentials. In this context,
a possible new scenario for a 
smooth exit from inflation to the radiation era is discussed by resorting the kination (stiff) era but without the inefficient 
radiation production mechanism of these models. This is done by considering rapidly decreasing expressions for $V(t)$ soon after inflation.
We show that the parameters of our models can reproduce the scalar
spectral parameter $n_s$ predicted by Planck data  
in particular for models with concave potentials. Finally, according to the recent BICEP2 data,
all our models allow a huge amount of 
primordial gravitational waves.
\end{abstract}
\section{Introduction}
Many experimental observations during the past decade (see \cite{1,q,2}) are in agreement with the hypothesis of a present day accelerating universe. In the standard 
$\Lambda$CDM cosmological model, an accelerating universe scenario invokes the presence of the so-called dark energy expressed in terms of a
cosmological constant $\Lambda$ representing about $70\%$ of the present universe matter-energy. However, the physical origin of this constant is still
obscure. Moreover, the physical mechanism leading to a small
$\Lambda$ that  begins to dominate only recently still remains mysterious. In view of the fact that a cosmological constant is the most simple solution of the Klein-Gordon equation for a scalar field $\phi$, it seemed reasonable to consider
a time-varying field,  named quintessence field, to describe a running cosmological constant driven by some potential $V(\phi)$
(see for example \cite{3,4,5,6}). This quintessence field produces a late times cosmological constant by means of a mechanism very closed 
to the one leading to primordial inflation. i.e. a slowly rolling scalar field 
(see for example \cite{7,8,9}). These models admit a tracker solution which partially alleviates
the coincidence problem. An unsolved problem in these models is fine-tuning, i.e the fact that
the energy density for $\Lambda$ is so small compared to typical particle
physics scales. Moreover, it is unclear how to obtain a smooth transition, after the inflationary epoch, to the radiation era, since the 
typical density of quintessence field is at least two orders smaller than the background density. 
Only in recent times the quintessence field should begin to dominate and thus
mimic a cosmological constant. As a result, what is practically absent in the literature of modern cosmology is an unified view in which 
a smooth transition from the inflationary epoch to the radiation era up to dark energy era is obtained.
An interesting alternative proposal to alleviate this lack
is to link primordial inflation to dark energy. Initially  
\cite{10,11,12}, this has been considered in the context of anthropic selection effects. More recently, in \cite{13} 
it was shown that primordial quantum fluctuations of an almost massless scalar field during the primordial inflation could explain the 
present quasi-de Sitter phase. Another interesting
paper along this line is \cite{14}. Here, the actual accelerating phase is 
obtained from primordial inflation by using the renormalization group equation to obtain the rate of change of the density of the 
vacuum energy as 
dictated by the usual approach of quantum field theory in curved spacetimes. The model also predicts
a transition to the radiation era.
However, a description in terms of an effective action together with the dynamics of $\phi$ is still missing.
In this paper, we follow this interesting line of research, but use a different approach.
In particular, we attempt to obtain a unified description of the whole history of the universe starting from the inflationary epoch by means of  phenomenological potentials, but initially expressed in term of the cosmic time $t$. 
In particular, we are interested in possible alternative mechanisms allowing a graceful exit from inflation 
together with a transition to the radiation era.
This paper is organized as follows. In section 2 we write down the relevant equations together with a presentation of our approach. 
In section 3 we study 
a particular model 
leading to the inflationary epoch.
In section 4 we analyze
the problem of a smooth exit from the inflation, while in section 5 we study the universal tracker de Sitter solution of our models.
In section 6 we study the dynamics of our model.
In section 7 we analyze our models in light of Planck data and present a study for a more general class of potentials allowing concave
potentials. In section 8 we study the possibility to introduce a running cosmological constant.
Finally, section 9 is devoted to some conclusions.
\section{Preliminaries}
To start with, we consider the line element appropriate for a Friedmann cosmology with constant spatial curvature $k$:
\begin{equation}
{ds}^2=-c^2{dt}^2+a^2(t)\frac{{dr}^2}{(1-kr^2)}+a^2(t)d{\Omega}^2.
\label{1}
\end{equation}
The relevant Einstein's field equations representing a cosmological perfect fluid with pressure $p(t)$ and energy density $\rho(t)$ 
together with an homogeneous and isotropic scalar field $\phi$ with pressure $p_{\phi}$ and energy density ${\rho}_{\phi}$
and with a potential $V(\phi)$ are:
\begin{eqnarray}
& &\frac{a_{,t}^2}{a^2}+k\frac{c^2}{a^2}=\frac{8}{3}\pi G\left(\rho+{\rho}_{\phi}\right),\label{2}\\
& &\frac{a_{,t,t}}{a}=-\frac{4\pi G}{3c^2}\left[c^2\rho+3p+c^2{\rho}_{\phi}+3p_{\phi}\right], \label{3}\\
& &\frac{{\phi}_{,t,t}}{c^2}+3H\frac{{\phi}_{,t}}{c^2}+\frac{dV}{d\phi}=0,\label{4}\\
& &{\rho}_{\phi}=\frac{{\phi}_{,t}^2}{2c^4}+\frac{V(\phi)}{c^2},\;\;\;p_{\phi}=\frac{{\phi}_{,t}^2}{2c^2}-V(\phi).\label{5}
\end{eqnarray}
In the usual approach to describe inflation one introduces a simple expression for $V(\phi)$. The most popular expression 
driving an inflationary primordial era is 
$V(\phi)\sim \frac{1}{2}m^2{\phi}^2$, though such potential is practically ruled out by the recent Planck data \cite{17}. 
However, the very recent BICEP2 data \cite{Bic} are in tension with Planck one concerning the value of the index $r$
measuring the primordial B-mode polarization(see section 6).\\
In any case, irrespective of the potential used,
the mechanism driving the exit from inflation up to the formation of usual matter
is not yet well established. 
To this purpose,
we take a different point of view by introducing 
'phenomenological' potentials in terms of the cosmic time $t$, i.e. agegraphic potentials.
Also in \cite{Chervon} the authors
consider particular expressions for 
$V(t)$ arising from a variational equation for the potential $\phi(t)$ leading to a particular form for $a(t)$. 
In particular, the expressions for $V(t)$ in \cite{Chervon}
allowing a transition to the usual Friedmann regime are the ones with
$V(t)\sim 1/t^2$. However, the transition occurs when $t\rightarrow\infty$, so it is not clear as to obtain a rapid transition in order to be in agreement with nucleosynthesis. As a result, we need of an expression for $V(t)$ quickly decaying after inflation. Moreover, the potential
$V(t)$ in order to satisfy Planck constraints must be nearly flat during inflation.
The aim of this paper is to explore explicit expressions for 
$V(t)$ that are in agreement with Planck and BICEP2 data 
allowing a graceful exit from inflation to the radiation era without invoking the usual reheating mechanism with an oscillating 
scalar field and avoiding the inefficient particle creation due to the gravitational field.
Our approach does not belong to the mainstream of the cosmological models, where a physically motivated potential
$V(\phi)$ is 
introducted. However, we point out
that by considering the potential $V(t)$ in terms of the cosmic time permit us to follow
directly the chronology of the universe.
Often in the literature the transition to the radiation dominated era is obtained asymptotically and it is not clear how to obtain a transition 
to the radiation era just before nucleosynthesis. Moreover, the recent Bicep2 experiment indicates that the inflation with a
single scalar field is again well posed but also that, thanks to the high running spectral index,  
the inflation can be well depicted by an effective expression of $V(t(\phi))$ that is not fixed in terms of $\phi$. As an example, at the 
beginning of the inflation, we can have an effective potential $V(\phi)\sim \phi$, 
i.e. the starting of the inflation is related to the axion monodromy potential, but 
after the de Sitter inflationary expansion we can have $V(\phi)\sim{\phi}^4$. Once a suitable expression for $V(t)$ is obtained,
we can reconstruct the usual expression $V(\phi)$ in the diferent cosmological epochs.
As we will show in this paper, a rapidly decreasing potential $V(t)$ allows a negligible fraction of radiation present soon after 
the inflation to dominate the universe.
For this purpose,
we are interested in expressions for $V(t)$ that are 
quickly decreasing soon after inflation. 
We set the potentials
\begin{equation}
V(t)=k_{\phi}+I_{\phi}e^{-\frac{{(t+kT)}^n}{T^n}}, n\geq 1
\label{ag1}
\end{equation}
where $k_{\phi}, I_{\phi}, T, k$ are  constant parameters with  $n$ a positive integer and $k\geq 0$ of the order of unity or less.
The form of the potential is not at all arbitrary: it must be chosen so that  
Planck constraints are satisfied, is nearly flat during the primordial inflationary era and with
a fast dacaying after inflation.  
The constant $k_{\phi}$ will be the future cosmological constant. Moreover, since we start with the inflationary epoch, 
we neglect the curvature term. Obviously, the idea to add a negligible constant or a quintessence potential
to the inflationary potential is not 
the novelty of this approach (see for example \cite{18} or \cite{19}). 
In practice,  the dark energy can be see as a small term $k_{\phi}$ in the classical potential
$V(t(\phi))$ 
generated by the primordial field survived up to dark matter era.\\ 
Concerning quantum fluctuations, as shown in \cite{13}, a quantum fluctuation  
can be frozen by the Hubble damping after inflation and thus produce a quintessence field.  
In \cite{19} a small constant relic
of primordial inflation is also considered.
Moreover,
in \cite{14}, by using renormalization group techniques, a negligible constant correction to the vacuum energy during inflation
becomes the actual cosmological constant. This shows that the idea that a small relic of early inflation becomes the actual cosmological 
constant is supported in the literature.
 
At this point, it is essential to stress that, in order to obtain physically reasonable models, we must look for monotonically decreasing 
expressions for $\phi$ in terms of $t$. In this way, we have a well defined expression for $V$ in terms of $\phi$, i.e. 
$V(\phi)=V(t(\phi))$. As a result, we can always regain the usual expression $V(\phi)$ which is important 
for a physical comprension of the inflationary mechanism.\\
In the next sections, we motivate the  choice (\ref{ag1}) together with a more general class of potentials discussed in section 6. 
\section{Inflationary era}
To start with, we firstly consider a particular simple class of  (\ref{ag1}) allowing simple computations: 
\begin{equation}
V(t)=k_{\phi}+I_{\phi}e^{-\frac{t}{T}},\;\;\;I_{\phi}>>k_{\phi}.
\label{6}
\end{equation}
Inflation is expected to begin at early times $t_i$ with 
$t_i \sim 10^{-36}\;s$ (in our setting $t=0$) and end to $t_e$ with $t_e\sim (10^{-33},10^{-32})\;s$.
We set to $t=0$ the starting of primordial inflationary era. 
Moreover, we take $V_{,\phi}=V_{,t}{({\phi}_{,t})}^{-1}$, that is certainly true
for ${\phi}$ monotonically decreasing. In this phase, the field equations are totally dominated by the scalar potential
$V(t(\phi))$.\\
To obtain the primordial inflationary epoch, the following usual conditions must hold:
\begin{equation}
\frac{|{\phi}_{,t,t}|}{c^2} <<\left\{3H(t)\frac{|{\phi}_{,t}|}{c^2},\;  |V_{,t}|  |{({\phi}_{,t})}^{-1}|\right\},\;\;
\frac{{\phi}_{,t}^2}{2c^2} << V(t).
\label{7}
\end{equation}
The parameter $T$ is expected to be of the order of the end of the inflationary epoch $t_e$ or greater.
Hence, the following condition holds during the quasi-de Sitter inflationary phase: 
\begin{equation}
\frac{t}{T}<1.
\label{8}
\end{equation}
Obviously, the approximation of this section will be more and more efficient provided that $t<<T$. In this case we have with good approximation a quasi de Sitter phase. Although in this section we perform the calculation with the potential 
(\ref{6}), similar computations hold for the more general potentials (\ref{ag1}).\\
As a title of example, by setting   $t/T\sim 1/10$ and $T\sim 10 t_e$
during inflation,  the end of the de Sitter inflationary epoch occurs when the potential is not more 
strictly constant and its variation is greater than $10$\%.\\
Under these assumptions, equation (\ref{4}) for the potential (\ref{6}) becomes:
\begin{equation}
{\phi}_{,t}^2=\frac{c^3I_{\phi}e^{-\frac{t}{T}}}{T\sqrt{24\pi G V(t)}}.
\label{9}
\end{equation}
In (\ref{9}) we have two possible roots.
For physical  purposes, we choose the one that is
monotonically decreasing (${\phi}_{,t}<0$) expression for $\phi$. Hence, 
we have a well defined expression for $V$ in terms of $\phi$, i.e. $V(\phi)=V(t(\phi))$.
By means of the condition (\ref{8}), after a Taylor expansion 
of (\ref{9}) we obtain:
\begin{eqnarray}
& &\phi={\phi}_0-At+Bt^2+o(t^2/T^2),\label{10}\\
& &A=\frac{\sqrt{2} 6^{\frac{3}{4}} c^{\frac{3}{2}}\sqrt{I_{\phi}}}{12\sqrt{T\sqrt{\pi G(k_{\phi}+I_{\phi})}}},\label{11}\\
& &B=\frac{\sqrt{6}(I_{\phi}+2k_{\phi})\sqrt{\sqrt{24}}\sqrt{I_{\phi}}c^{\frac{3}{2}}}{96T(k_{\phi}+I_{\phi})\sqrt{T\sqrt{\pi G(k_{\phi}+I_{\phi})}}},
\label{12}
\end{eqnarray}
where ${\phi}_0$ is the initial value for ${\phi}$.
It is easy to see that inequalities (\ref{7}) are satisfied for:
\begin{equation}
T>> T_I\simeq \frac{c}{10\sqrt{\pi G(k_{\phi}+I_{\phi})}}.
\label{13}
\end{equation}
Inequality (\ref{13}) will be studied in the next section.
In this way, our particular model can reproduce inflation with an Hubble flow $H_I$ given by:
\begin{equation}
H_I\simeq \frac{1}{c}\sqrt{\frac{8}{3}\pi G(k_{\phi}+I_{\phi})}.
\label{14}
\end{equation}
As a consequence, inequality (\ref{13}) becomes:
\begin{equation}
T>>T_I\simeq \frac{\sqrt{8}}{10\sqrt{3}H_I}.
\label{14b}
\end{equation}
Finally, by denoting with $N$ the e-folds number, we obtain the approximate formula:
\begin{equation}
t_e \geq \frac{N}{H_I}.
\label{15}
\end{equation}
For the reasoning above, 
$t_e$ denotes the end of the phase where $H_I$ is practically constant. However, the accelerated expansion lasts for a few small time
after $t_e$. Hence, to solve the flatness problem, $N$ in (\ref{15}) can be less than $60$.
It should be noticed that 
all the physical conditions above are independent of the initial value of $\phi$, i.e ${\phi}_0$. This is 
interesting since fine-tuning for ${\phi}_0$ is avoided.
Since in our approach we use a kind of inverse method, to a better understanding of the inflationary mechanism, we need to
reconstruct the potential in the usual field theory form, i.e. $V(\phi)$. In the case under consideration, and also for the generalizations of 
section 6, what is relevant during inflation is the behaviour (\ref{10}) up to the first term  with respect to $t$, i.e.
$\phi={\phi}_0-At+o(1)$. The term $\sim t^2/T^2$ is relevant for the calculation of the spectral index $\eta_V$ (see section 6).
After solving with respect to the cosmic time we obtain, during the initial (de Sitter) inflationary epoch, the following Taylor expansion:
\begin{equation}
V(\phi)=V_0+ \frac{I_{\phi}}{TA}\left(\phi-{\phi}_0\right)+o(\phi-{\phi}_0),\;\;V_0=k_{\phi}+I_{\phi}.
\label{ax}
\end{equation}
The effective part of the reconstructed potential, at the beginning of the inflation
($t\simeq 0$), the potential $V(t(\phi))$ is linear with respect to $\phi$ with excellent approximation.
The leading term in (\ref{ax})
recalls the one of 
axion monodromy potentials but with higher order corrections. These corrections become more and more relevant when approaching the end of the inflation.  
In this way we have a time varying expression for $V$ in terms of $\phi$, while in the usual
approach this behaviour is fixed.   
Moreover, as shown in this section,
the initial value ${\phi}_0$ is irrelevant for the inflationary mechanism to work. Hence we could set ${\phi}_0=0$. Otherwise,
we can choose 
${\phi}_0=(ATV_0)/I_{\phi}\sim AT$ and thus we regain the usual expression for the 
potential in axion monodromy models for chaotic inflation at $t\simeq 0$.
\section{From inflation up to recombination}
Also in this section and in section 5 we use for practical computations  the simple potential (\ref{6}), but the conclusions also apply to the
more general one (\ref{ag1}).
First of all, $k_{\phi}$ must be identified with the actual cosmological constant, i.e.
$k_{\phi}=\Lambda c^4/(8\pi G)$. Concerning the ratio between the inflationary
cosmological constant ${\Lambda}_I\sim I_{\phi}$
and the value of $\Lambda$ , we 
are unable to reliably calculate the expected vacuum energy. In practice, only
theoretical estimates of various contributions to the vacuum energy
density in quantum field theory are at our disposal. From these estimations we have 
that ${\Lambda}_I/\Lambda \sim (10^{40}, 10^{120})$.
Since $H_I^2\sim I_{\phi}$ and 
$H_0^2\simeq k_{\phi}\sim\Lambda$,  
we have ${\Lambda}_I/\Lambda=I_{\phi}/k_{\phi}\sim (10^{40}, 10^{120})$ and
$H_I/H_0\sim (10^{20},10^{60})$. 
Moreover, conditions (\ref{14}) and (\ref{14b})   
imply that $T>> t_P$, being $t_P$ the Planck time. Hence, 
for the ratio $I_{\phi}/k_{\phi}$ the range  $I_{\phi}/k_{\phi}\geq \sim 10^{100}$ seems most viable since a smaller value would 
imply a rather unrealistic value for $T$. In fact, it is expected that the scalar field $\phi$ drives the inflationary epoch at early times 
and not when 
radiation  forms,  before nucleosynthesis. In the range  $I_{\phi}/k_{\phi}\geq \sim 10^{100}$, also a small time of the order of
$T\sim 10^{-32}s,10^{-31}s$ works. In practice, $T$ can be chosen of the order of  $(1,10^2)t_e$.
With these estimates, as an example,
for  $t>\sim 10^{-28} s$ we have that $V(t)= k_{\phi}(1+o(1))$. This means that the potential (\ref{6}) (and (\ref{ag1}) obviously), 
after the time $t\sim t_e$ rapidly converges to the value $k_{\phi}$. 
However, this does not imply that $\phi$ becomes rapidly constant, i.e
our models converge rapidly to the standard cosmological constant scenario. 
This happens if and only if during inflation, in addiction to the conditions
(\ref{7}), we have ${\phi}_{,t}^2\sim c^2 k_{\phi}$. But this situation is not interesting since we have not a way to obtain a smooth transition 
from inflationary era to the usual matter era in a natural simple way.
The opposite condition during inflation
(i.e.  ${\phi}_{,t}^2>> c^2 k_{\phi}$, but $<< c^2 V(t)$), 
is obtained for  $T<< (10^{50}, 10^{60}) t_0$, with $t_0$ the actual age of the universe. Obviously this 
condition is satisfied in our case. Hence, we have a scenario such that also after the inflation $\phi$ is
monotonically decreasing, but  much less rapidly with respect to  $V(t(\phi))$. As a result, after a very short transition epoch
($t\sim 10^{-28} s$ by using  $T\sim 10^{-31} s$) the relevant equation for $\phi$ becomes:
\begin{equation}
{\phi}_{,t}\left[{\phi}_{,t,t}+3H{\phi}_{,t}\right]=0+o(1).
\label{17}
\end{equation}
Hence, shortly after the inflation, it is justified to consider the following
expansion for $\phi$: $\phi={\phi}_0(1+o(1))$, where ${\phi}_o$ is solution of the homogeneous
part of equation (\ref{17}) and the terms $o(1)\sim I_{\phi}/Te^{-t/T}$ (for the potential (\ref{6})) and are exponentially small. 
By setting ${\phi}_{,t}^2=\psi$, the equation (\ref{17}), after the inflation and before that the term $k_{\phi}$ begins to dominates with respect to the 
kinetic one ${\phi}_{,t}^2$ in $H(t)$ (i.e. at the late times de Sitter phase),
becomes:
\begin{equation}
\frac{1}{2}{\psi}_{,t}+\frac{{\psi}^{\frac{3}{2}}}{c^2}\sqrt{12\pi G}  = 0.
\label{18}
\end{equation}
Note that
during inflation, when $V$ dominated on the term ${\phi}_{,t}$, equation (\ref{17}) is homogeneus
with respect to
$\psi(t)$ with the second member given by 
$I_{\phi}/Te^{-t/T}$ by using (\ref{6}). Hence,  the field $\phi$ started to receive contributions from the homogeneous
part plus the particular solution of the non homogeneous one. These inhomogeneous contributions must be zero
as $T\rightarrow 0$. This fact also motivates the reasonings leading to equation (\ref{18}), soon after inflation.\\
The solution for (\ref{18}) look as follows: 
$\phi=B-c^2/(\sqrt{12\pi G})\ln[(t/c^2)\sqrt{12\pi G}-C]$, 
where $B,C$ are integration constants.
For our purposes and without loss of generality we can set $C=0$
(this can be done for example by a simple constant shift in the time.). 
Hence we have that 
the density of the scalar field ${\rho}_{\phi}$ at the stiff era is 
${\rho}_{\phi}=1/(24\pi G t^2)$, i.e. it scales as ordinary matter in a power low dominated cosmology
with $a(t)\sim t^{\frac{1}{3}}$. Note that with this solution, the conditions 
${\phi}_{,t}^2>>c^2\{2k_{\phi},V(t(\phi))\}$ and
$H{\phi}_{,t}^2>>c^2|V_{,t}|,\;|{\phi}_{,t}{\phi}_{,t,t}|>>c^2|V_{,t}|$ 
are satisfied at early epochs after inflation up to dark matter era
and garantee the correctness of the approximation leading to equation (\ref{18}).\\
Thanks to the condition chosen, this solution is equivalent to ordinary matter with a stiff equation of state,
i.e. ${p}_{\phi}= c^2{\rho}_{\phi}$ and with the typical behaviour
$\{{\rho}_{\phi}, p_{\phi}\}\sim 1/t^2$ of ordinary hot big bang cosmology.  
Note that the equation of state 
$p_{\phi}=c^2{\gamma}_{\phi}{\rho}_{\phi}$
for the scalar field, just before  the stiff dominated era
(leading to the approximate equation (\ref{18})) and after the pure inflationary epoch (${\gamma}_{\phi}\simeq -1$), 
runs very quickly from the value ${\gamma}_{\phi}\simeq -1$ soon after the inflation up to the stiff value 
${\gamma}_{\phi}\simeq 1$. This implies that  in a very short time, the factor ${\gamma}_{\phi}$  assumed all the values between
$[-1,1]$ and as a result ordinary matter and radiation has been created during this short transition phase. Hence, 
a possible scenario is that a fraction    
of radiation created  soon after inflation ($t\in(T, 10^3 T)$), or just present during inflation
but negligible with respect to ${\rho}_{\phi}$,
was in thermal equilibrium when a quasi-stable stiff era (${\rho}_{\phi \sim 1/t^2}$) is reached.\\
In particular, during this post
inflationary  epoch, 
we can have the decoupling $\phi\rightarrow \phi_1+\chi$, 
where $\chi$ is a light field such that $|\phi_1|>>|\chi|$ and 
${\phi}_{,t}^2>>{\chi}_{,t}^2$. Another viable possibility is that a negligible amount of radiation, 
depicted by $\chi$, has been present soon after primordial inflation together with the inflationary potential $\phi$.\\
In \cite{24} has been shown that light scalar fields can drive the last stage of inflation by means of dominant 
quantum fluctuations proportional to some positive power of ${\phi}_0$
and eventually create a 
new inflationary phase. Fortunately, these arguments are not applicable in our context for several reasons. First of all,  in our scenario
the light field 
$\chi$ comes into action when the equation of state for ${\rho}_{\phi}, p_{\phi}$ is not inflationary, i.e. ${\gamma}_{\phi}\geq -1/3$.
Moreover, the initial value ${\phi}_0$ is inessential to begin inflation and thus can be set to zero in our framework.\\
As a result, after the 'stiff' era, we may have a  universe  composed of two
perfect fluids: an ordinary one that follows the standard hot big bang cosmology (${\chi}$) and a 
scalar field component (${\phi}_1$) with a stiff equation of state
decaying more rapidly with respect to ordinary matter
up to dominate the universe a late epoch with a de Sitter inflationary epoch.\\ 
After this 'decoupling', the ordinary matter decays as $\sim 1/t^2$, while ${\rho}_1$ and $p_1$ decay following 
equation (\ref{17}), but now with $H(t)\sim \alpha /t$.\\
Note that reheating  scenarios without oscillations of $\phi$ 
can be traced back to (for example) the papers \cite{20,21,22}. In particular, the mechanism to create matter with a non oscillating field, whose kinetic part is much greater than the potential $V(\phi)$, has been named 'kination'
in \cite{22}. In this picture, the conventional 
reheating mechanism is substituted by quantum field effects in curved spacetimes similar to the Hawking 
radiation formation, provided that a huge cosmological expansion $H(t)$ arises.
A criticism to this approach can be found in  \cite{23},  where a non oscillating scalar field $\phi$ can
suffer for a lack of efficient mechanisms to describe the reheating phase after the inflation, exception made for the preheating scenarios, although for sufficient huge values for $H(t)$ this mechanism could be efficient too (see \cite{22}).\\
However, the mechanism described above concerning our models  can be eventually added to the 'curvaton'
mechanism during kination. In our approach, the parameter $T$ permit us to fix the  time transition to the stiff era. Hence,
in the scenario we have proposed, a fraction of radiation (together with other kinds of particles)
created during the transition to the stiff era
(${\phi}_{1,t}^2\sim c^2 V(t(\phi_1))$) or just present during inflation, can be thermalized during the stiff era well before
nucleosynthesis. After this decoupling, the density ${\rho}_1$ looks as ${\rho}_1\sim a^{-6}(t)$, while the radiation after the decoupling
of $\phi$ follows the usual big bang scenario.
To be more quantitative, consider the time evolution of the ordinary radiation energy
${\rho}_r$ during the stiff era:
\begin{equation}
{\rho}_{r,t}+3H{\rho}_r(1+{\gamma}_r)=0,\;\;{\gamma}_r=\frac{1}{3},\;\;\;H=\frac{1}{3t}.
\label{dar1}
\end{equation}
As a result, by denoting with ${\rho}_{ir}$ the initial fraction of the radiation (created just before the stiff era) at the beginning of the 
stiff era $t=t_i$ we have for (\ref{dar1}) the solution ${\rho}_{r}={\rho}_{ir}{(t_i/t)}^{4/3}$. For ${\rho}_{\phi}$ during the stiff era we have
(see reasonings below equation (\ref{18}))
${\rho}_{\phi}=1/(24\pi G t^2)$. Hence, for the ratio  ${\rho}_{\phi}/{\rho}_r$ we obtain
\begin{equation}
\frac{{\rho}_{\phi}}{{\rho}_r}=\frac{{\rho}_{i\phi}}{{\rho}_{ir}}
{\left(\frac{t_i}{t}\right)}^{\frac{2}{3}}.
\label{dar2}
\end{equation}
Equation (\ref{dar2}) shows that by setting ${\rho}_{i\phi}>>{\rho}_{ir}$, after a short time well before nucleosynthesis, the
radiation becomes dominant with respect to stiff matter. As an example, by setting $t_i=10^{-27}s$ and 
${\rho}_{i\phi}\sim 10^3{\rho}_{ir}$, at the time $t\sim 10^3 t_i$, radiation becomes the $\sim 10\%$ of the whole matter content of the
universe. Moreover,
for ${\rho}_{i\phi}\sim 10^5{\rho}_{ir}$ at $t\sim 10^6 t_i$ ($\sim 10^{-21}s$), we have again that radiation becomes
the $\sim 10\%$ of the stiff matter. 
As a consequence, a negligible fraction of radiation created soon before the stiff era 
(or present during inflation) 
can dominates the universe by a very short time.  
Hence, stiff era quickly ends and radiation era with $H=1/2t$ starts well before nucleosynthesis.
Also note that
a fraction of baryonic matter can survive at the stiff era with the same mechanism depicted above. These simple computations
show that the depicted mechanism allowing a smooth exit from inflation can be physically viable with the potentials
(\ref{ag1}).
At the recombination up to low redshifts of the order of some units, we have, from 
(\ref{17}),  ${\rho}_{1}\sim k^2/t^4$, with $k$ an integration constant.
By equation (\ref{2}), we obtain the leading  correction to the behaviour of
$H$, i.e. $H(t)=\alpha/t+\beta/t^3+o(1), \beta>0$ both for (\ref{6})
(\ref{ag1}). For the scale factor $a(t)$ we obtain, after integrating (\ref{2}),
$a(t)\sim t^{\frac{2}{3}} e^{-\frac{\beta}{2t^2}}+o(1)$. These corrections are small 
ensuring certainly a small deviation  from the standard
$\Lambda$CDM scenario at least from recombination epoch. 
Hence we expect agreement with the usual constraints (Supernovae Type Ia, CMB..). After recombination
${\phi}_{1,t}^2/c^2$ becomes so small to be comparable with $k_{\phi}$, and as a consequence the cosmological constant born at
primordial inflation begins to dominate. It should be noted that, thanks to these corrections, after recombination our models 
are  not
coincident with the concordance one. More consistent deviations can be obtained by considering a time variable term instead of
$k_{\phi}$ in (\ref{6}).
We stress that the scenario discussed above is only a possibility. More precise quantum calculations are essential to take this 
hypothetical scenario as sound.
In any case,
We believe that  the interesting physical fact is the existence of early times, shortly after inflation, 
such that $\{{\rho}_{\phi}, p_{\phi}\}\sim 1/t^2$ allowing a creation in a short period of radiation and baryonic matter,  
since this can allow a possible smooth transition to the usual hot big bang scenario.
The important mathematical fact is that for a quickly decaying potential the equation for $\phi$ becomes homogeneous, and as a result soon after inflation $\phi$ has a power law expression.
Another realistic possibility is that a certain amount of radiation can be created soon before inflation due to black hole evaporation
(see for example \cite{scar}) and dominate the universe after inflation with the mechanism depicted above.
\section{Late times de Sitter era}
The important condition that must be satisfied in order to have a late time tracker
 de Sitter era is:
\begin{equation}
V_{,t} = o({\phi}_{,t}),\;\;\;for\;\;\;t\rightarrow \infty.
\label{19}
\end{equation}
Under this condition, the leading terms at $t\rightarrow\infty$ for equation (\ref{4}) by using (\ref{6}) is:
\begin{equation}
{\phi}_{,t,t}+\frac{1}{c}\sqrt{24\pi G k_{\phi}}\; {\phi}_{,t}=0,
\label{20}
\end{equation}
with solution
\begin{equation}
\phi=B_1+B_2e^{-\frac{1}{c}\sqrt{24\pi G k_{\phi}} t},
\label{21}
\end{equation}
with $B_1,B_2$ two arbitrary constants. Condition (\ref{19}) is fulfilled for
\begin{equation}
T<\frac{1}{c\sqrt{3\Lambda}}.
\label{22}
\end{equation}
From equation (\ref{14b}) we see that (\ref{22}) is fulfilled if and only if $c\sqrt{\Lambda}<<H_I$. 
Hence, thanks to the previous section estimates and to the present day  very small value of the cosmological
constant,  
condition (\ref{22}) is obviously verified and a 
de Sitter inflationary phase emerges independently on the physical parameters of the model. This means that 
a late times de Sitter phase is a tracker solution for our model.
Finally, note that also at the present time $t_0$ the term $k_{\phi}$ in the potential (\ref{ag1}) can dominate on
the kinetic one, 
provided that the rather reasonable assumption on $B_2>0$ is imposed:
\begin{equation}
B_2<<\frac{c^2 e^{\left(c/2\sqrt{3\Lambda}\;t_0\right)}}{\sqrt{12\pi G}}.
\label{dar3}
\end{equation}
In fact, we look for a monotonically decreasing $\phi$,  and certainly condition (\ref{dar3}) can be satisfied for $t<<t_0$ after 
recombination.
Only when
the dust density of the universe ${\rho}_d$ is of the order of ${\rho}_d\sim k_{\phi}/c^2$, the universe starts to accelerate.
\section{Reconstruction of the potential and dynamics of the scalar field}
For sake of simplicity in this section
we study the features of our potential (\ref{6}), but similar arguments follow for the one 
(\ref{ag1}). To start with, thanks to (\ref{2}) and (\ref{4}) and after setting ${\phi}_{,t}^2/c^2=K(t)$
we have:
\begin{equation} 
\frac{K_{,t}}{2}+K\sqrt{24\pi G\left({\rho}_{m}+\frac{K}{2c^2}+\frac{V}{c^2} \right)}=\frac{I_{\phi}}{T}e^{-\frac{t}{T}},
\label{d1}
\end{equation}
where ${\rho}_m$ denotes the usual matter-energy content of the universe. For the reasonings of the section $4$, we may suppose that this quantity is completely negligible up to the kination era. Starting from the kination epoch, also a small fraction of radiation can 
quickly dominate the matter-energy content of the universe. Hence, during inflation up to kination we can set 
${\rho}_m=0$.\\ 
During the primordial inflation, with the slow roll approximation we have the expansions (\ref{10}) and (\ref{ax}). Note that in this approximation, the equation (\ref{d1}) can be integrated. The resultant solution is rather cumbersome, but the main
properties of the primordial de Sitter era are depicted by the following effective expression for the potential $V(\phi)$:
\begin{equation}  
V(\phi)=k_{\phi}+I_{\phi}+\frac{I_{\phi}}{AT}(\phi-{\phi}_0)+I_{\phi}
\left[\frac{B}{TA^3}+\frac{1}{2A^2T^2} \right]{\left(\phi-{\phi}_0\right)}^2+o(1).
\label{d2}
\end{equation}
As pointed in section 3, at the starting of the early inflationary epoch, the effective potential (\ref{d2}) is practically linear in the 
field $\phi$. During this fase, the kinetic term $K(t)$ is monotonically decreasing (see the expansion (\ref{10}) and below)
and $K(t)<<V(t), K(t)>>k_{\phi}$.
At the end of inflation, when ${\phi}_{,t}^2/c^2\sim V(t)$,
the behaviour of $V(\phi)$ becomes more complicated with respect to the behaviour (\ref{d2}), but $K$ is again monotonically
decreasing with $V(t)$ dropping rapidly to the value $k_{\phi}$.
It is important to note that also at the kination era, i.e. when ${\phi}_{,t}^2/c^2>>V(t)$, equation (\ref{d1}) implies that
$K_{,t}<0$. Since the right hand side of (\ref{d1}) is certainly negligible from the kination era, 
this implies that from inflation up to late times the kinetic term is always monotonically decreasing.\\
At the kination era 
($K>>V, {\rho}_m<<K/c^2, t\geq t_{ik}$), equation (\ref{d1}) can be integrated and the potential can be put in the form:
\begin{equation}
V(\phi)=k_{\phi}+I_{\phi}e^{-\frac{t_{ik}}{T}Q(\phi)},\;\;\;Q(\phi)=e^{-\frac{\sqrt{12\pi G}}{c^2}(\phi-{\phi}_{ik})}.
\label{d3}
\end{equation}
Note that to obtain expression (\ref{d3}),  for sake of simplicity, 
an integration constant, namely $K_{ik}$ has been used to have $K\sim 1/t^2$ during kination
era.
Also note that the potential (\ref{d3}) is rapidly converging to $k_{\phi}$. In fact 
$\phi-{\phi}_{ik}\leq 0$ and tipically (see the reasonings at section 4) $t_{ik}/T\geq \sim 1000$. Hence soon after the begin of the 
kination era, the small radiation present in ${\rho}_r$ begins to dominate on $K$ and radiation era begins 
well before nucleosynthesis. For completeness, we can give the expression fo $t(\phi)$ during the radiation era:
\begin{equation}  
V(\phi)=k_{\phi}+I_{\phi} e^{-\frac{t(\phi)}{T}},\;\;\;t(\phi)=
\frac{1}{{\left[\frac{\phi-{\phi}_{ir}}{2c\sqrt{K_{ir}}t_{ir}^{\frac{3}{2}}}+\frac{1}{t_{ir}^{\frac{1}{2}}}\right]}^2}.
\label{dd1}
\end{equation} 
Starting from the radiation era and up to decoupling, the potential $V$ plays no role in the universe dynamics, that is well depicted by the usual hot big bang scenario. However, a non vanishing $K$ is always present leading to 
corrections of the usual scenario that could be detectable. Also in this case we give the expression 
for $t(\phi)$ during the dark matter era, after decoupling:
\begin{equation}  
V(\phi)=k_{\phi}+I_{\phi} e^{-\frac{t(\phi)}{T}},\;\;\;t(\phi)=
\frac{1}{\frac{\phi-{\phi}_{im}}{c\sqrt{K_{im}}t_{im}^2}+\frac{1}{t_{im}}}.
\label{dd2}
\end{equation} 
This situation changes when 
$K/c^2\sim{\rho}_m<k_{\phi}/c^2\simeq V(\phi)/c^2$. There, a new inflation arises leading to the current acceleration of the 
universe. The potential during the late times inflationary epoch is given by (see section 5):
\begin{equation}  
V(\phi)=k_{\phi}+I_{\phi}{\left(\frac{\phi-{\phi}_{\infty}}{B_2}\right)}^{\frac{c}{T\sqrt{24\pi Gk_{\phi}}}},
\;\phi-{\phi}_{\infty}>0,\;B_2>0,\;(\phi-{\phi}_{\infty})/B_2<1. 
\label{d4}
\end{equation}
The term depending on $\phi$ in (\ref{d4}) is completely negligible with respect to $k_{\phi}$.
In this way, we have reconstructed the expression for $V(\phi)$ at the cosmological era: during inflation 
(formula (\ref{d2})), during kination (formula (\ref{d3}), during radiation era (formula (\ref{dd1})), during dark matter era
(formula (\ref{dd2})) and finally at the late times de Sitter era (formula (\ref{d4})).\\
Note that,  similarly to the potential proposed in \cite{18},
we can also obtain an approximate (but qualitatively equivalent to the one with $V$ expressed as a time function)
description of our model with $V$ given in terms of $\phi$. To this purpose, we must match the expressions 
(\ref{d2})-(\ref{d4}) by imposing the continuity of $V$ together with the first derivative of $V(\phi)$, i.e.
$V_{,\phi}(\phi)$ at the various cosmological era. This is done in the appendix where, to obtain a better reconstruction,
a transition zone between the end of primordial inflation and the begin of kination is introduced.
Summarizing, the important ingredients are: the monotonically decreasing behaviour for $K(t)$; the rapidly converging
behaviour for $V(t)$ to $k_{\phi}$ soon after inflation; the fact that $K(t)$ decreases after inflation much less quickly 
with respect to $V(t)$.\\  
Equations (\ref{10})-(\ref{12}) show that during inflation up to $t\leq T$, $K$ is monotonically decreasing, but the ratio
$V/K$ remains practically constant during thid time. For $t>T$, the potential $V(t)$ rapidly converges to $k_{\phi}$ and then the second member of (\ref{d1}) becomes rapidly converging to $0$ and then $K$ contuinues to be monotonically decreasing for
all times. As a numerical example, consider $T=10^{-33} s$ and a huge inflation near the Planck scale with 
$I_{\phi}\sim 10^{107}Kg/(ms^2)$. During the inflation up to $t\sim T$, we have that
$V\simeq I_{\phi}$. After a time $t\simeq 3.22\;10^2 T$, we have that $V(t)$ becomes of the order of unity ($1 Kg/ms^2$).
After a time $t\simeq 3.32\;10^2 T$, we have practically $V(t)\simeq k_{\phi}$. Hence, after a very short time after 
$T$, the behaviour of $K$ is completely detrminated by the left hand side of (\ref{d1}), while the role of $V\simeq k_{\phi}$ on the dynamics of the universe is completely negligible. Hence, $K$ assumes a power law behaviour 
(see the discussion after equation (\ref{18})).
Only after recombination (see section 5), when
$K<k_{\phi}$ a new inflationary era emerges.                 
To appreciate this rapid variation of the dynamic soon after $t=T$,
consider equation (\ref{d1}) with $K=q(t)V$. Typically, at the begin of the inflation we have that 
$q_0\in[10^{-10}, 10^{-3}]$, it depending on the value of $I_{\phi}$. Since $K$ is chosen monotonically decreasing, from 
(\ref{d1}) we have 
\begin{equation}  
q\sqrt{1+\frac{q}{2}}>q_0 F(t),\;\;\;F(t)=\frac{I_{\phi}e^{-\frac{t}{T}}}{V(t)}
\;\sqrt{\frac{k_{\phi}+I_{\phi}}{V(t)}}.
\label{d5}
\end{equation}
Note that during the inflationary epoch up to $t\simeq T$, the function $q(t)$ is practically constant and also 
$F(t)$ is, with $F(t)\simeq 1$ with obviously $q<<1$. The situation quickly changes for $t>T$. As a numerical example,
consider again $T=10^{-33}s$. Up to $t\leq T$, $F(t)$ is of the order of the unity 
($F(t)\simeq 1.7$ for $t=T$). But for $t=10T$ we have $F(t)\simeq 148$, while for $t=10^2 T$ we found 
$F(t)\sim e^{50}$. As a consequence,  in a very short time after $t=T$, $q(t)$ enormously increases from values
$q<<1$ up to very huge values and at $t=10^2 T$ the kinetic term $K$ 
becomes at least $\sim q_0 e^{100/3}$ times greater than $V$ and kination begins.
These examples  are sufficient to illustrate the fate of $K$ in the rapid transition from inflation to kination era. As shown in the discussion after equation (\ref{dar2}), a negligible fraction of radiation at the kination era quickly leads to a radiation era well
before nucleosynthesis (as well known radiation expands much more rapidly with respect to stiff matter).
From the reasonings above it is evident that the delicate point is the transition 
from inflation to kination. To this purpose, we have numerically integrated equation (\ref{d1}) from inflation up to kination. 
In fig. 1 we have considered the cases with $T=10^{-33}s$, $T=10^{-32}s$ and $I_{\phi}/k_{\phi}=10^{117}$,
$I_{\phi}/k_{\phi}=10^{104}$. The time is expressed in Planck units. From fig. 1 we see that in both cases when 
$t\simeq 10 T$ we have $K=V$. After this time, the potential $V$ very quickly converges to $k_{\phi}$ and 
$K>>V$ and as a consequence kination era begins soon after $t\simeq 10T$. At the kination $V\simeq k_{\phi}$ and the reasonings above take place. The integration confirms the essential fact that $K$ (and $\phi$) are monotonically decreasing for all times since 
starting from the kination era the right hand side of equation (\ref{d1}) is completely negligible. In fig. 2  we have plotted 
the ration $V/K$.
\begin{figure}[h!]{\psfig{file=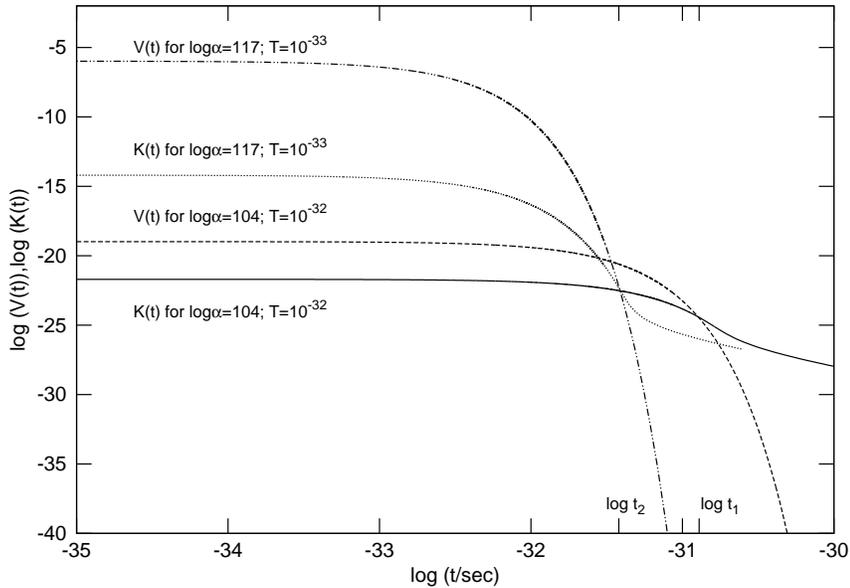,width=8cm, angle=-90} 
        \caption{In figure the behaviour of $log V(t)$ and $log K(t)$ vs. $log(t/sec)$ ($t$ is the time). The figure shows plots corresponding to the following choices of parameters: ($\alpha=10^{104}$; $T=10^{-32}$ sec) and ($\alpha=10^{117}$; $T=10^{-33}$ sec). It also shows the corresponding times $log(t_1/sec)=-30.89$ and $log(t_2/sec)=-31.42$ for which $V(t) = K(t)$ for both the choices of parameters.}
	\label{fig2}}
\end{figure}
\begin{figure}[h!]{\psfig{file=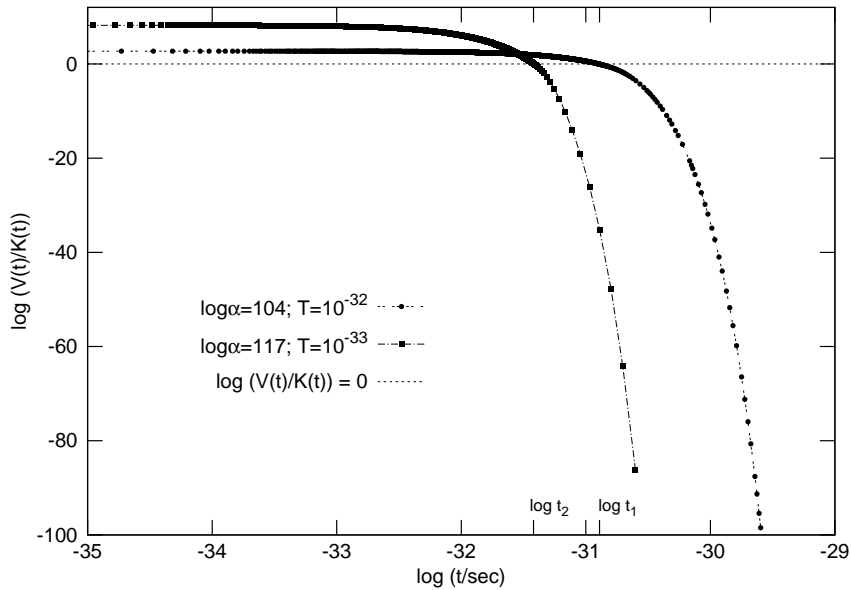,width=8cm, angle=-90} 
        \caption{$log (V(t)/K(t)$ vs. $log(t/sec)$ ($t$ is the time). The figure shows two plots of $log (V(t)/K(t))$ corresponding to the following choices of parameters: ($\alpha=10^{104}$; $T=10^{-32}$ sec) and ($\alpha=10^{117}$; $T=10^{-33}$ sec). It also shows the plot $y=0$ and the corresponding times $log(t_1/sec)=-30.89$ and $log(t_2/sec)=-31.42$ for which $log (V(t)/K(t))=0$ for both the choices of parameters.}
	\label{fig2}}
\end{figure}
\section{Planck and BICEP2 data}    
The recent Planck results  \cite{17}
posed serious constraints to the inflationary scenario, ruling out popular models 
($V\sim {\phi}^p,p\geq 2$). The  particular model given by (\ref{6}) 
predicts a small positive value 
for the slow-roll parameter ${\eta}_V\simeq 3/(4TH_I)$, while Planck data favor potentials with
$V_{,\phi,\phi}<0$.
Nevertheless, our particular model can be tested against Planck constraints for running spectral index $n_s$.
For the scalar spectrum parameter $n_s$ and 
the tensor spectral index $r$ during the beginning of the inflation we obtain:
\begin{equation}
n_s\simeq 1-\frac{3}{2TH_I},\;\;{\epsilon}_V\simeq \frac{1}{2TH_I},\;\;
r\simeq 16{\epsilon}_V.
\label{Pla1}
\end{equation}
The Planck constraint, i.e. $n_s\in[0.955,0.967]$, can be satisfied in the running case, where 
$r<0.26$ (otherwise $r<0.11$), but with a value for $n_s\sim 0.97$.
By setting $T\simeq  10^{-32} s$, $I_{\phi}/k_{\phi}\sim 10^{102}$ we have 
$r\simeq 0.16$. More precisely, we have to set $1/(TH_I)\simeq 0.02$.
The same result can be obtained by setting $T\sim 10^{-33} s$ and $I_{\phi}/k_{\phi}\sim 10^{104}$. 
Moreover, all the conditions raised in the paper are satisfied since $T/ T_I\simeq 320$ (see (\ref{14b}))
and with $t_e\in \sim (10^{-34},10^{-32})s$. 
The  particular model (\ref{6}) predicts a huge amount of gravitational waves. This is at odd for Planck, but is good for BICEP2.
The BICEP2 result \cite{Bic} strongly suggest a running scalar spectrum index $n_s$ with
$r$ in the range
$r \in[0.15,0.27]$ with the central value $r=0.2$. Note that, by setting $1/(TH_I)=0.025$, we obtain $r=0.2$ and 
$n_s=0.9625$, in perfect agreement with the value of Planck esperiment. If Bicep2 data where confirmed, 
then our simple model is able to 
allow a correct amount of  scalar to tensor perturbations together with the central value for $n_s$ predicted by Planck.
 
To obtain a negative value for $\eta_V$ together with a smaller amount of gravitational waves within Planck data, we must consider the potential (\ref{ag1}) with a suitable $k>0$ and $n>1$.
As a title of example, we consider explicitally the case with $n=3$.
During the de Sitter inflationary phase, for $t<<T$, we can  thus neglect $k_{\phi}$  with respect to $I_{\phi}$. We have:
\begin{equation}
{\phi}_{,t}\simeq -\frac{k\sqrt{2}}{4e^{\frac{k^3}{4}}}c^{\frac{3}{2}}
\sqrt{\frac{\sqrt{24}\;I_{\phi}}{T\sqrt{\pi G I_{\phi}}}},\;\;
{\phi}_{,t,t}\simeq \frac{(-4+3k^3)\sqrt{2}}{16 e^{\frac{k^3}{4}}}c^{\frac{3}{2}}\sqrt{\frac{\sqrt{24}\;I_{\phi}}{T^3\sqrt{\pi GI_{\phi}}}}.
\label{ag2}
\end{equation}
The conditions (\ref{7}) are satisdied by a condition similar to (\ref{13}); i.e. with $k\in(0,1)$ we have 
\begin{equation}
T>>\frac{(4-3k^3)\sqrt{8}}{4k\sqrt{3}H_I}.
\label{ag3}
\end{equation}
During inflation, in the de Sitter like  phase ($t<<T$),  By denoting $H_I^2\simeq 8\pi G I_{\phi}/(3c^2)$, 
for the indices we thus obtain:
\begin{eqnarray}
& &\epsilon_V\simeq\frac{3k^2 e^{\frac{k^3}{2}}}{2T H_I},\nonumber\\
& &\eta_V\simeq\frac{e^{\frac{k^3}{2}}}{4kT H_I}\left(9k^3-4\right),\nonumber\\
& &n_s\simeq 1-\frac{e^{\frac{k^3}{2}}}{TH_I}\left[\frac{9k^2}{2}+\frac{2}{k}\right].
\label{ag4}
\end{eqnarray}
From (\ref{ag4}) we see that $\eta_V<0$
if and only if $k<{(4/9)}^{1/3}$. As a title of example, for $k=1/2$ we have, by setting $e^{1/16}/(TH_I)\simeq 0.0074$ (which is 
always possible for the reasonings
below equation (\ref{Pla1})), $\epsilon_V\simeq 0.0028, r\simeq 0.04, \eta_V\simeq -0.01$,  and as
a result we obtain the best fit Planck value
$n_s\simeq 0.962$. 
Generally,  in order to satisfy the Planck constraints,
for fixed $T$ and $n$ ($>1$), a smaller value for k implies  
a greater value for $H_I$. To this purpose,
note that $k$ cannot be arbitrary small. 
The transition to the stiff era also arises for (\ref{ag1}) in a very short time after the early inflationary epoch
one given by (\ref{6}) for $n>1$.
Also in this case, according to BICEP2, we can easily obtain a value for $r\simeq 0.2$.
Also for the potentials (\ref{ag1}), we can reconstruct the effective potential during the initial phase of inflation ($t/T<<1$).
Once again, we obtain as a dominant
effective part during these very early times
a behaviour
similar to (\ref{ax}) with $V(t(\phi))\sim\phi+const$. In particular, for $t \simeq 0$, the potential $V(t(\phi))$
with good approximation is linear with respect to $\phi$.
Hence,
our inflationary potentials can mimic axion monodromy models at the beginning of the inflation suggesting that the
starting mechanism for inflation could be in our case related to the one of these models motivated by chaotic inflation.
In particular, this is a general properties for exponentially time decaying potentials given by (\ref{ag1}),
with an expansion during inflation at $t\simeq 0$ given by:
\begin{equation}  
V(\phi)=V_0+V_1(\phi-{\phi}_0)+V_2{(\phi-{\phi}_0)}^2+higher\;\;orders,\;\;\;V_1,V_2\in R.
\label{axg}
\end{equation}
With the flowing of the time during inflation,  the nonlinear terms neghligible at $t\simeq 0$, become more and more relevant.
These examples show how with the present approach we can obtain a good agreement with Planck 
and BICEP data and at the same time
a possible simple mechanism explaining the matter formation in the early universe together with a unification between
dark energy and an early de Sitter phase by a single potential. 
In particular, the existence of stiff matter in the early universe can be useful to explain
the abundance of heavy elements in our universe.
\section{Building late times running dark energy}
Concerning the issue related to the small value of the cosmological constant, it can be alleviated by introducing
a running cosmological constant in (\ref{ag1}.)
As an example, in (\ref{ag1}) we could take
$k_{\phi}\rightarrow k_{\phi}(t/\tau)^n e^{-\frac{t}{\tau}}, \tau\sim t_0,\;n\geq 0$, where $t_0$ is the present day cosmic time.
In this way we obtain a late times 'quintessence'-like field with 
an effective cosmological constant running to zero at late times. Also in this case, 
the physics up to the recombination is practically left unchanged with respect to the one given by (\ref{ag1}). 
However, a more physically motivated
possibility, which seems the most natural in our context, is the one given in terms of the so called
agegraphic dark energy \cite{hol1,hol2}. This kind of holographic dark energy is motivated by applying arguments of quantum mechanics and
general relativity. The energy density is given by:
\begin{equation}  
{\rho}_q=\frac{3F^2}{8\pi G t^2},
\label{ageo1}
\end{equation}  
where the positive constant $F$ ($F>1$ to have  a current accelerated universe) 
takes into account quantum effects in curved spacetimes. Unfortunately, the expression (\ref{ageo1}) 
cannot have a matter-dominated phase if $F <1$ and without an interaction with the dark matter, as shown in
\cite{hol3}. In \cite{hol3} this problem has been solved by considering the conformal time $\eta$ instead of $t$. However, in our context,
it seems more natural to solve this issue by considering, as suggested in \cite{hol3}, the potential:
\begin{equation}
V(t)=\frac{3F^2 c^2}{8\pi G {(t+\tau)}^2}
+I_{\phi}e^{-\frac{{(t+kT)}^n}{T^n}}, n\geq 1\;\;\;F>1,
\label{agage}
\end{equation}
where the cut-off $\tau$ is greater than the recombination time and less than the time where cosmic structures form.
In this manner, the agegraphic term is certainly negligible with respect to the exponential one in the early inflationary phase, and also in the 
kination and the radiation era since there ${\Omega}_q<<1$.
Hence, all the reasonings of the sections 2,3,4 and formulas (\ref{ag4}), (\ref{axg})
are left unchanged. For the late times de Sitter phase, the agegraphic dark energy can mimic
the cosmological constant (see \cite{hol3}). As correctly argued in \cite{hol3}, the expression (\ref{ageo1}) has been originary
motivated by a 'gedanken experimenten' concerning distances measurement with light-clocks in Minkowski 
spacetime. However, in the early universe the cosmological expansion affects the possible localizing measure
(see \cite{TV,TV2} in the context of non commutative geometry). 
Only when the value of the cosmic flow is sufficiently small, the Minkoskian 
limits can be obtained,
as shown in \cite{TV2}.  As a consequence the radiation dominated scenario discussed in section $4$
remains valid and the 
problem of the small value of $k_{\phi}$ is alleviated. Moreover, note that it is not clear as to obtain a primordial inflationary epoch 
together with the
transition to the radiation era 
in the usual ageographic context, i.e. without the exponential term in 
(\ref{agage}). To be more quantitative, from (\ref{agage}),
the condition such that the agegraphic term is completely negligible during primordial inflation is 
\begin{equation}  
\tau >>\frac{F}{H_I},\;t<\tau,
\label{inc} 
\end{equation}  
in practice $\tau>>t_e$.
Moreover, we must have  a small Hubble flow in order to have a physical motivation for the 
Minkowskian expression (\ref{ageo1}). 
To this purpose, it is interesting
to note that, as shown in \cite{TV2}, for a flat Friedmann
spacetime, rather less limitations arise for the uncertainty of a 
coordinate with respect to the Minkowskian case due to the role of a huge Hubble flow $H$.
Another possibility to take into account the role of the Hubble flow according to the reasoning above is to take a time variable 
$F(t)$:
\begin{equation}
V(t)=\frac{3F^2(t) c^2}{8\pi G {(t+t_i)}^2}
+I_{\phi}e^{-\frac{{(t+kT)}^n}{T^n}}, n\geq 1\;\;\;F> 0
\label{agage2}
\end{equation}
where the time $t_i$ is a time of the order of the begin of the primordial inflation in the usual timing of the primordial inflation. 
This time is due to the fact that in this paper the start of the inflation is fixed at $t=0$.\\
Hence, the condition (\ref{inc}) during the primordial inflation becomes:
\begin{equation}  
F(t)<<t_i H_I,\;\;\;t\leq t_e,
\label{inc2} 
\end{equation}  
In order to allow the dark matter to dominate, after recombination
the function $F(t)$ must be chosen of the order of unity.
After the recombination, the role of the Hubble flow in gedanken 
measurements becomes negligible and Minkowskian expression (\ref{ageo1}) is regained. 
It should be stressed that 
the new agegraphic potentials expressed in terms of the conformal time $\eta$ (see \cite{hol2,hol3})
are 'phenomenological'. In fact, the use of a conformal time instead of the age of the universe $t$ in (\ref{ageo1}) is motivated 
by phenomenological purposes. Since $\eta=\int dt/a(t)$,
the original derivation for the 
metric fluctuations in Minkowski spacetime due to the presence of clock devices can be done in terms of the conformal time $\eta$, but its expression is not obtained by merely substituing $t$ with $\eta$ in (\ref{ageo1}).  
Hence, the naive use of the expression (\ref{ageo1}) with $t\rightarrow \eta$ loses its original motivation. In our context, both the terms
$k_{\phi}$ and the one of this section can be see as representing small fluctuations during the primordial inflation survived up to dominate
the recently the history of the universe.
Summarizing, with the potentials (\ref{ag1}), (\ref{agage}) and
(\ref{agage2}), we have a well defined de Sitter inflationary era motivated by axion monodromy potential,
together with a simple mechanism for the transition to the radiation era up to a physically motivated term leading to 
the late times de Sitter phase.  
\section{Conclusions and final remarks}
This paper is an attempt to obtain a unified description of the universe from the inflationary era up to the late 
de Sitter phase. In particular, we introduce a physically viable
class of phenomenological potentials $V(t)$ that allow us to achieve this unified description. 
In fact, our models are able to produce an inflationary mechanism both at early and late times. We show that
a smooth transition from the inflationary epoch up to the radiation era can be obtained. 
After recombination, the scalar field $\phi$ generates small corrections to the Hubble flow $H$ that look as 
$H\sim \frac{2}{3t}+\frac{const.}{t^3}+o(1)$. Although these corrections are expected
 very small, they could be investigated by future cosmological data.
Moreover, a smooth transition to the de Sitter accelerate phase is obtained.\\ 
The models (\ref{ag1})
have four arbitrary constants, i.e. $k_{\phi}, I_{\phi}, T, k$. Two of them
can be fixed by data concerning the present value of the cosmological constant ($\sim k_{\phi}$) and the
effective cosmological constant driving the primordial inflation ($\sim I_{\phi}$). 
Unfortunately,  
we have not at our disposal a sound estimate for this ratio from ordinary 
quantum field theory.
The other parameters can be chosen in light of Planck and BICEP2 data.\\
The parameter $T$ represents the 
characteristic time after which primordial inflationary 
de Sitter-like expansion phase is not more efficient. The ones of the section $7$ have $F(t)$ instead of $k_{\phi}$.
 
We have also tested our class of models with the recent Planck and BICEP2 data, in particular for the 
indices $n_s, \epsilon_V, r, \eta_V$. 
We found that the model with a convex potential can be acceptable if we consider the running spectral index
Planck constraints and the more recent
BICEP2 data. By considering more general potentials (\ref{ag1})with $n>1$, we can easily obtain 
concave potentials with an acceptable range ($<0.12$) for the tensor ratio $r$ according with Planck data.
More general potentials can be used provided that $V(t)$ be 
monotonically decreasing with respect $t$, nearly constant during inflation and rapidly decreasing soon after inflation to a small 
value, i.e. the cosmological constant, or to a small running cosmological constant, as shown in section $7$.
However, irrespective of the modifications on $V(t)$, our study suggests that the definition of $V$ as an agegraphic potential
allows to introduce potentials with a time varying form in terms of $\phi$. In fact, at the begin of the primordial inflation
our potentials (\ref{ag1}) and (\ref{agage2}) have with good approximation a linear expression in terms of $\phi$. 
There, a well posed mechanism to start inflation in terms of axion monodromy
is given,  further motivating the choices (\ref{agage2}) and (\ref{ag1}). 
However,  also during inflation other higher orders do appear. This is in line with the idea that quantum fluctuations can add further terms
to the initial expression for $V(\phi)$ due to renormalization group equation (to this purpose see \cite{14}). Moreover, 
the explicit presence of $t$ permit us to  follows the timing of the universe just in time to obtain the fundamental transition to the 
radiation era before nucleosynthesis an up to the dark matter era. Often in the literature  these transitions are obtained only 
asymptotically (see \cite{Chervon}), which is unphysical.
In our approach we consider from the onset
a simple physically motivated expression 
for $V$ in terms of the cosmic time $t$, provided that a monotonically decreasing expression
for $\phi$ is allowed. For this reason we named the agegraphic potential 
'phenomenological'.  The exponential form of the potentials (\ref{6}) and (\ref{ag1}) is dictated from the necessity to have an efficient
primordial inflationary mechanism together with a rapidly decaying allowing a plausible simple mechanism for a 
graceful exit from inflation up to matter creation 
and to a late times de Sitter phase. In this regard, note that an arbitrary naive choice for $V(t)$ generally does not work. As an extreme 
example, since the class of potentials (\ref{ag1}) and (\ref{agage2})
rapidly converge to a constant or negligible value, one may be tempted to 
set a potential of the form $V(t)=I_{\phi}$ for $t<T$ and $V(t)=k_{\phi}$ for $t>T$. Apart from a lack of  continuity at
$t=T$ that does not permit a smooth transition from the two regimes at $t<T$ and $t>T$, these models predict a strictly gaussian perturbation,
i.e. $\eta_V=\epsilon_V=0, n_s=1$, in complete disagreement  with Planck data, which are the cornerstone of any physically viable 
cosmological model.
 
In this manner we have a possible unification of the two inflationary epochs by a single smooth agegraphic potential
expressed in terms of the cosmic time $t$ in such a way that dark energy is 
a relic of primordial inflation 
together with a possible simple graceful 
exit from inflation up to radiation era without invoking the usual reheating mechanism or dangerous particles creation due to the gravitational field.
It is also interesting to note that in this framework the mechanism chosen for primordial inflation 
is not plagued from
fine-tuning of ${\phi}_0$ since the only condition that leads to an efficient inflationary mechanism
is  that the characteristic time $T$ is of the order or some order larger than  $t_e$, i.e. the time 
after which inflation is not more strictly in a de Sitter phase. Theoretically, we could take
${\phi}_0=0$ and inflation works as well. 
Obviously, the mechanism leading up to radiation era obviously must be further investigated by quantum calculations, 
by resorting the post inflationary non oscillatong scenarios, but read in our frame and with a different mechanism for matter-radiation 
creation without  encounter the drawbacks of usual non-reheating cosmologies (see \cite{23}). To this purpose note that the mechanism 
proposed leading to nucleosynthesis is not plagued from fine tuning problems and only requires a negligible remnant of radiation 
soon after inflation. 
\section*{Appendix}
In this section we obtain an effective description of our potential $V$ in terms of the variable $\phi$. To this purpose, we adopt the 
expressions (\ref{d2})-(\ref{d4}) to put $V(\phi)$ in a form similar (but more complicated)
to the one present in \cite{18}  with the domain expressed in terms 
of $\phi$. The calculations are simple but the final expressions for the relations among the parameters
are rather cumbersome. Hence in this appendix we only present the  potential together with the 
suitable equations for the mactching conditions.\\
We denote with ${\phi}_{ik}, {\phi}_{ir}, {\phi}_{im}, {\phi}_{s}$ the values of the potential respectively at the begin of
kination, radiation, dark matter era and structures formation. We have added another transition zone, indicated with
${\phi}_{tr}$,  that describes the short transition between the early inflationary epoch and the kination era. 
We have:
\begin{eqnarray}
V(\phi)&=&k_{\phi}+I_{\phi}+\frac{I_{\phi}}{AT}(\phi-{\phi}_0)+\nonumber\\
&+& I_{\phi}
\left[\frac{B}{TA^3}+\frac{1}{2A^2T^2} \right]{\left(\phi-{\phi}_0\right)}^2,\;\;\phi\in[{\phi}_{tr},{\phi}_0];\label{A1}\\
&=& k_{\phi}+I_{\phi} e^{-\frac{t(\phi)}{T}},\;\;t(\phi)=-a \phi+b,
\;\;\phi\in[{\phi}_{ik}, {\phi}_{tr}] \label{AA1}\\
&=& k_{\phi}+I_{\phi}e^{-\frac{t(\phi)}{T}},\label{A2}\\
& & t(\phi)=\frac{c^2\beta}{\sqrt{12\pi G}}\left[\left(1+\frac{\sqrt{12\pi G}t_{ik}}{c^2\beta}\right)
e^{-(\phi-{\phi}_{ik})\frac{\sqrt{12\pi G}}{c^2}}-1\right],\nonumber\\
& &\beta=\frac{1}{c\sqrt{K_{ik}}}-\frac{\sqrt{12\pi G}}{c^2}t_{ik},\;\;\phi\in[{\phi}_{ir},{\phi}_{ik}],\nonumber\\
&=& k_{\phi}+I_{\phi} e^{-\frac{t(\phi)}{T}},\;\;t(\phi)=
\frac{1}{{\left[\frac{\phi-{\phi}_{ir}}{2c\sqrt{K_{ir}}t_{ir}^{\frac{3}{2}}}+\frac{1}{t_{ir}^{\frac{1}{2}}}\right]}^2},\;\;
\phi\in[{\phi}_{im}, {\phi}_{ir}];\label{A3}\\
& =& k_{\phi}+I_{\phi} e^{-\frac{t(\phi)}{T}},\;\;\;t(\phi)=
\frac{1}{\frac{\phi-{\phi}_{im}}{c\sqrt{K_{im}}t_{im}^2}+\frac{1}{t_{im}}},\;\;
\phi\in[{\phi}_s, {\phi}_{im}];\label{A4}\\
&=&k_{\phi}+I_{\phi}{\left(\frac{\phi-{\phi}_{\infty}}{B_2}\right)}^{\frac{c}{T\sqrt{24\pi Gk_{\phi}}}},\;\;
\phi\in[{\phi}_{\infty}, {\phi}_s].\label{A5}
\end{eqnarray}
The next step is to impose the matching conditions at 
$\phi={\phi}_{tr},\phi={\phi}_{ik}, \phi={\phi}_{ir},\phi={\phi}_{im},\phi={\phi}_s$. To obtain a 
continuous and differentiable expression for $V(\phi)$, we must impose the continuity of $V(\phi)$ and $V_{,\phi}(\phi)$
at that points. To start with, we must paste the functions (\ref{A1}) and (\ref{AA1}) at $\phi={\phi}_{tr}$. From the continuity  
of $V(\phi)$ and its derivative we obtain: 
\begin{eqnarray} 
& & I_{\phi}+\frac{I_{\phi}}{AT}({\phi}_{tr}-{\phi}_0)+ I_{\phi}
\left[\frac{B}{TA^3}+\frac{1}{2A^2T^2} \right]{\left({\phi}_{tr}-{\phi}_0\right)}^2=I_{\phi}e^{-\frac{t_{tr}}{T}},\label{A6}\\
& &\frac{I_{\phi}}{AT}+2 I_{\phi} \left[\frac{B}{TA^3}+\frac{1}{2A^2T^2} \right]({\phi}_{tr}-{\phi}_0)=
\frac{I_{\phi}}{T}e^{-\frac{t_{tr}}{T}}a,\label{A7}
\end{eqnarray} 
where
\begin{equation} 
t_{tr}\simeq 10^{-2} T=-a{\phi}_{tr}+b.
\label{AA2}
\end{equation}
The equations (\ref{A6})-(\ref{AA2}) permit to obtain the parameters ${\phi}_{tr},a,b$.\\
Concerning the transition at ${\phi}_{ik}$ we must paste (\ref{AA1}) with (\ref{A2}) at $\phi={\phi}_{ik}$. We obtain:
\begin{eqnarray}
& & t_{ik}\simeq 10^3 T =-a{\phi}_{ik}+b, \label{AA3} \\
& & \beta+\frac{\sqrt{12\pi G}\;t_{ik}}{c^2}=a. \label{AA4}
\end{eqnarray}
From (\ref{AA3}) we can obtain ${\phi}_{ik}$ (with the condition ${\phi}_{ik}<{\phi}_{tr}<{\phi}_0$). With this value we can obtain
$\beta$ from (\ref{AA4}).\\
The same calculations must be done for the functions (\ref{A2}) and
(\ref{A3}) at $\phi={\phi}_{ir}$. The continuity is obtained by calculating the function $t(\phi)$ in (\ref{A2}) at
$\phi={\phi}_{ir}$ and substituing the expression so obtained for $t({\phi}_{ir})=t_{ir}$
in the expression for $t(\phi)$ given by (\ref{A3}).  For the differentiability condition we have:
\begin{equation} 
\left(\beta +\frac{\sqrt{12\pi G} t_{ik}}{c^2}\right)e^{-({\phi}_{ir}-{\phi}_{ik})}=\frac{1}{c\sqrt{K_{ir}}},
\label{A8}
\end{equation} 
that can be easily solved for $K_{ir}$. Concerning the matching between the potentials (\ref{A3}) and (\ref{A4}),  also in this case the continuity at $\phi={\phi}_{im}$ is obtained by calculating $t(\phi)$ in (\ref{A3}) at $\phi={\phi}_{im}$ and putting the obtained expression 
in the formula for $t(\phi)$  in (\ref{A4}). For the differentiability we have:
\begin{equation}  
t({\phi}_{im})=
\frac{1}{{\left[\frac{{\phi}_{im}-{\phi}_{ir}}{2c\sqrt{K_{ir}}t_{ir}^{\frac{3}{2}}}+\frac{1}{t_{ir}^{\frac{1}{2}}}\right]}^3}
\frac{1}{t_{ir}^{\frac{3}{2}}\sqrt{K_{ir}}}=\frac{1}{\sqrt{K_{im}}}.
\label{A9}
\end{equation}
From equation (\ref{A9}) we can easily obtain $K_{im}$. Finally, we can paste the potentials (\ref{A4}) and
(\ref{A5}) when dark energy era begins to start, i.e. at the epoch of the formation of the big structures in the universe 
(at redshift $z\simeq 0.5$). The continuity equation it gives:
\begin{equation}
t(\phi)=
\frac{1}{\frac{{\phi}_s-{\phi}_{im}}{c\sqrt{K_{im}}t_{im}^2}+\frac{1}{t_{im}}}=-
\frac{c}{\sqrt{24\pi G}}\ln\left(\frac{{\phi}_s-{\phi}_{\infty}}{B_2}\right).
\label{A10}
\end{equation} 
Equation (\ref{A10}) can be solved analitically in terms of $B_2$. The differentiability condition at $\phi={\phi}_s$ can be solved
analitically in terms of ${\phi}_{\infty}$. Summarizing, in this way we have the parameters
${\phi}_0, {\phi}_s$ left free for the potential $V(\phi)$.
The potential so obtained can generate the physics depicted in this paper. 
The traduction of $V(t)$ in terms of $\phi$ is important since it permit us to study the model of this paper as an ordinary
field theory.\\
In fig. 3 we have plotted the behaviour of $V(\phi)$ (vs $-\phi$)
for $T=10^{-33}$ and ${\phi}_0=0$ (remember that in our tractation 
${\phi}_0$ plays no role) in Planck units.  The reconstructed potential, as stated above, is a monotonic increasing function of
$\phi$ (decreasing with respect to $-\phi$). Note that the potential shows two plateau concerning the begin of the primordial inflation
($\phi=0$) and the begin of the kination era up to the begin of the late times de Sitter phase ($\phi={\phi}_{\infty}$).
\begin{figure}[h!]{\psfig{file=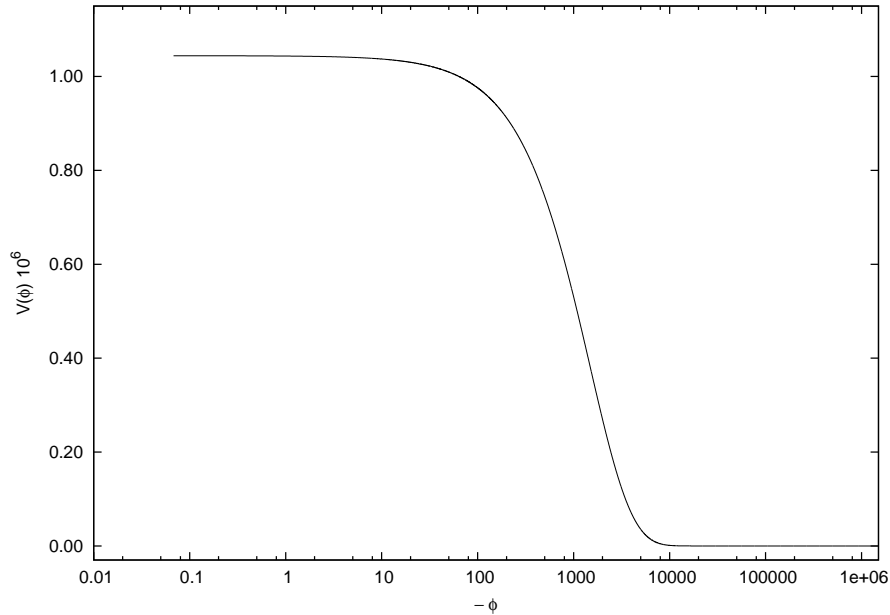,width=12cm, angle=0} 
        \caption{$V({\phi})$ vs. $-{\phi}$ (${\phi}$ is negative) to have a logaritmic scale. $V({\phi})$ has been multiplied by $10^6$ to have a more readable y range. The figure shows the plot of $V({\phi})$ for the following choice of parameters: ($\alpha=10^{117}$; $T=10^{-33}$ sec). It is apparent the practically constant value of the $V({\phi})$ in the range $\phi\in[{\phi}_{tr},0]$, the rapidly decrease in the range $\phi\in[{\phi}_{ik}, {\phi}_{tr}]$ and again the practically constant behaviour with $V\simeq k_{\phi}$
        in the range $\phi\in[{\phi}_{\infty}, {\phi}_{ik}]$} 
	\label{fig3}}
\end{figure}

\end{document}